\documentclass[graybox]{svmult}

\usepackage{mathptmx}       
\usepackage{helvet}         
\usepackage{courier}        
\usepackage{type1cm}        
\usepackage{makeidx}         
\usepackage{graphicx}        
\usepackage{multicol}        
\usepackage[bottom]{footmisc}

\makeindex             

\begin{document}

\title*{Uncovering the network structure of the world currency market:
Cross-correlations in the fluctuations of daily exchange rates}
\titlerunning{World currency market correlation network}
\author{Sitabhra Sinha and Uday Kovur}
\institute{Sitabhra Sinha \at The Institute of Mathematical Sciences, 
CIT Campus, Taramani, Chennai 600113, India.\\ \email{sitabhra@imsc.res.in}
\and Uday Kovur \at Department of Physics, Birla Institute of Technology \&
Science, Pilani 333031, India.}
%
%
\maketitle

\abstract*{The cross-correlations between the exchange rate
fluctuations of 74 currencies over the period 1995-2012 are analyzed
in this paper. The eigenvalue distribution of the cross-correlation
matrix exhibits a bulk which approximately matches the bounds
predicted from random matrices constructed using mutually uncorrelated
time-series. However, a few large eigenvalues deviating from the bulk
contain important information about the global market mode as well as
important clusters of strongly interacting currencies. We reconstruct
the network structure of the world currency market by using two
different graph representation techniques, after filtering out the
effects of global or market-wide signals on the one hand and random
effects on the other. The two networks reveal complementary insights
about the major motive forces of the global economy, including the
identification of a group of potentially fast growing economies
whose development trajectory may affect the global economy in the
future as profoundly as the rise of India and China has affected it in
the past decades.
}

\abstract{
The cross-correlations between the exchange rate 
fluctuations of 74 currencies over the period 1995-2012 are analyzed 
in this paper. The eigenvalue distribution of the cross-correlation 
matrix exhibits a bulk which approximately matches the bounds 
predicted from random matrices constructed using mutually uncorrelated 
time-series. However, a few large eigenvalues deviating from the bulk
contain important information about the global market mode as well as
important clusters of strongly interacting currencies. We reconstruct
the network structure of the world currency market by using two 
different graph representation techniques, after filtering out the 
effects of global or market-wide signals on the one hand and random 
effects on the other. The two networks reveal complementary insights 
about the major motive forces of the global economy, including the
identification of a group of potentially fast growing economies
whose development trajectory may affect the global economy in the
future as profoundly as the rise of India and China has affected it in
the past decades.
}

\section{Introduction}
At whatever scale one studies economic phenomena, we can find complex systems,
comprising relatively large number of mutually interacting elements 
often connected to each other in non-trivial topologies, at work.
The components can be individual traders, firms, banks, markets or countries,
but however complicated the behavior of the individual agents in the system,
an even richer collective behavior is manifested 
at the scale of the entire group of interacting agents. Explaining the emergence of 
such systems-level phenomena which may be qualitatively different from the
properties exhibited by the individual components is one of the key
goals of many physicists working on socio-economic questions, an enterprise
that is often referred to as {\em econophysics}~\cite{Sinha11}. An
important step in this direction will be to identify features of economic
systems that are {\em universal}, in the sense of occurring at many different
scales, suggesting that their existence is not contingent upon the particular
conditions prevailing in a specific situation. This will help econophysicists
to focus on phenomena that are not just the outcome of a series of
historical accidents and which can therefore be potentially explained by 
generalizable mechanisms.

Market dynamics has been identified by many physicists as a
particular area of economics that has the potential for yielding
several such universal features. In particular, one can mention the
identification of scale-invariant distributions in price fluctuations,
the trading volume and number of trades~\cite{Gabaix03,Farmer05} in
equities markets (but see also Ref.~\cite{Vikram11a}). However, in
order to get an understanding of how qualitatively new features emerge at the
level of the collective dynamics of the entire market, one needs to
understand the nature and structure of interactions between the
agents. While several studies on the networks underlying equities
markets (e.g., Ref.~\cite{Sinha07}) have been done, we need to compare
between markets of different kinds in order to distinguish those
features that are particular to specific systems and those which are
universal. With this aim, we undertake a detailed investigation of the
world currency market in this article. While several previous studies
have looked at the cross-correlations between the foreign exchange
rates of different currencies
(e.g., see Refs.~\cite{Ausloos01,Mizuno06,Drozdz07}), our results
reveal several novel insights and unexpected features of the network of
interactions between the currencies that we reconstruct from the
cross-correlations data. The period of the preceding sixteen years 
we have chosen for our study has seen remarkable transformations in
the world economy with the emergence of new economic powerhouses such
as China and India, but it has also shown how our world is vulnerable
to massive system-spanning crises (such as that of 2007-08). The study
of networks in the global currency market provides an important
perspective with which to view the positive as well negative impacts of
globalization. It has been argued that globalization is neither a
completely new phenomenon in world history nor are its effects always beneficial to the
economy~\cite{Jennings11}. We hope that by investigating
the collective dynamics of the international trade in currencies in
order to identify the major motive forces of the world economy, one
can potentially understand the long-term trends and prospects
of globalization.

\section{The World Currency Market}
The foreign exchange (FX) market, representing the entire global
decentralized trading of various currencies, is the largest financial
market in the world with an average daily trading volume estimated
in 2010 to be $4 \times 10^{12}$ US Dollars~\cite{BIS10}.
A typical trade in the FX market consists of a pair of agents
exchanging a
certain amount of a particular currency for a mutually agreed amount
of another currency. The ratio of the amounts of the two currencies
changing hands specify the corresponding {\em exchange rate} for the pair
of currencies concerned. Thus the exchange rates determine the value
of a currency with respect to another (the numeraire). The modern FX
market characterized by a large number of currencies having floating
exchange rates which continuously fluctuate over time date from the
1970s. The varying rates reflect the changing demand and supply
for the currencies, and are thought to be directly influenced by the
trade deficit/surplus of the corresponding
countries~\cite{Sarkar06} as well as macroeconomic variables such as
changes in growth of the gross domestic product, interest rates, etc.
However, international events can often trigger large perturbations in
the FX market and it is possible that sudden changes in the exchange
rates of a certain group of currencies can spread over time,
eventually affecting a much larger number of currencies.
Our article aims at uncovering the network of interactions between
the different currencies of the FX market along which perturbations can propagate
in the world currency market.

{\em Description of the data set.} We have considered the daily exchange
rate of currencies in terms of US Dollars (i.e., the base currency) 
publicly available
from the website of the financial services provider company, Oanda
Corporation~\cite{oanda}. We have chosen the US Dollar as the
numeraire as it is currently the primary reserve currency of the
world and is most widely used in international transactions.
The daily rates are computed as the average of all exchange rates
(taken as the midpoint of the bid and ask rates) quoted during a
24-hour period 
prior to the day of posting the rate.
For cross-correlation analysis, we have focused on the price data of $N = 74$
currencies from October 23, 1995 to April 30, 2012, which corresponds
to $T = 6034$ working days. 
The choice of currencies was governed by our decision to only include
those which either follow a free float or a managed float exchange
rate regime. We have thus avoided currencies such as the Chinese yuan
whose rate of exchange is pegged against another currency so that the
value of currency does not vary appreciably in time (resulting
in trivial cross-correlations). We have also
excluded countries having a dollarized economy such as Panama, Ecuador
Vietnam or Zimbabwe, that use a foreign currency - in majority of
cases, the US
Dollar - instead of or alongside the domestic currency, as this
introduces strong artifacts in the cross-correlations.
The period of observation was chosen so as to maximize the volume of
available data.
Using the MSCI Market Classification Framework~\cite{MSCI}
we have divided the countries to which the currencies belong into
three categories: developed, emerging and frontier markets. This
classification is based on a number of criteria including market
accessibility, size and liquidity of the market and the sustainability
of economic development. While many of the OECD countries belong to
the developed category, the rapidly growing economies of Asia, Africa
and Latin America (such as the BRICS group comprising Brazil, Russia,
India, China and South Africa) are in the emerging category with the
frontier markets category being populated by the remainder.
The individual currencies, along with the above economic classification
of the corresponding countries and the geographical regions to which they belong are
given in Table~\ref{ss:table1}.

\begin{table} [tbp]
\centering
\caption{The list of 74 currencies analyzed in this article arranged according to type of
market and grouped by geographical region.}

\begin{tabular}{clrrr}\hline\hline
 $ $ &  &  &  &  \\ 
 $i$ & Currency code & Currency name  & ~Type of market & Geographical region \\ 
 \hline
   1 & CAD   & Canadian Dollar &  Developed & Americas            \\  
   2 & DKK    & Danish Krone &  Developed & Europe and Middle-East   \\  
   3 & EUR & Euro & Developed & Europe and Middle-East     \\  
   4 & ILS  & Israeli New Shekel &  Developed & Europe and Middle East           \\  
   5 & ISK  & Iceland Krona &  Developed & Europe and Middle-East          \\  
   6 & NOK & Norwegian Kroner  &  Developed & Europe and Middle-East           \\  
   7 & SEK & Swedish Krona &  Developed & Europe and Middle-East        \\  
   8 & CHF    & Swiss Franc &  Developed & Europe and Middle-East      \\  
   9 & GBP  & Great Britain Pound &  Developed & Europe and Middle-East    \\  
  10 & AUD & Australian Dollar &  Developed & Asia-Pacific          \\  
  11 & HKD & Hong Kong Dollar &  Developed & Asia-Pacific  \\
  12 & JPY  & Japanese Yen &  Developed & Asia-Pacific  \\
  13 & NZD & New Zealand Dollar &  Developed & Asia-Pacific  \\
  14 & SGD  & Singapore Dollar &  Developed & Asia-Pacific  \\
  15 & BOB & Bolivian Boliviano &  Emerging & Americas  \\
  16 & BRL  & Brazilian Real &  Emerging & Americas    \\
  17 & CLP  & Chilean Peso &  Emerging & Americas    \\
  18 & COP & Colombian Peso & Emerging & Americas    \\
  19 & DOP &  Dominican Republic Peso    & Emerging & Americas   \\
  20 & MXN  & Mexican Peso &  Emerging & Americas    \\
  21 & PEN & Peruvian Nuevo Sol  & Emerging & Americas    \\
  22 & VEB  & Venezuelan Bolivar  &  Emerging & Americas    \\
  23 & ALL   & Albanian Lek  &  Emerging & Europe, Middle-East and Africa  \\
  24 & DZD  & Algerian Dinar  &  Emerging & Europe, Middle-East and Africa  \\
  25 & CVE & Cape Verde Escudo &  Emerging & Europe, Middle-East and Africa   \\
  26 & CZK & Czech Koruna &  Emerging & Europe, Middle-East and Africa   \\
  27 & EGP & Egyptian Pound  &  Emerging & Europe, Middle-East and Africa   \\
  28 & ETB   & Ethiopian Birr &  Emerging & Europe, Middle-East and Africa   \\
  29 & HUF  & Hungarian Forint &  Emerging & Europe, Middle-East and Africa   \\
  30 & MUR & Mauritius Rupee    &  Emerging & Europe, Middle-East and Africa   \\
  31 & MAD   & Moroccan Dirham   &  Emerging & Europe, Middle-East and Africa    \\
  32 & PLN & Polish Zloty  &  Emerging & Europe, Middle-East and Africa   \\
  33 & RUB & Russian Rouble &  Emerging & Europe, Middle-East and Africa   \\
  34 & ZAR & South African Rand   &  Emerging & Europe, Middle-East and Africa    \\
  35 & TZS     & Tanzanian Shilling   &  Emerging & Europe, Middle-East and Africa    \\
  36 & TRY  & Turkish Lira    &  Emerging & Europe, Middle-East and Africa    \\
  37 & INR & Indian Rupee      &  Emerging & Asia  \\
  38 & IDR  & Indonesian Rupiah   &  Emerging & Asia   \\
  39 & KRW   & South Korean Won             &  Emerging & Asia    \\
  40 & PHP & Philippine Peso     &  Emerging & Asia    \\
  41 & PGK  & Papua New Guinea Kina      &  Emerging & Asia    \\
  42 & TWD & Taiwan Dollar        &  Emerging & Asia      \\
  43 & THB   & Thai Baht      &  Emerging & Asia \\
  44 & GTQ & Guatemalan Quetzal       & Frontier & Americas \\
  45 & HNL    & Honduran Lempira  & Frontier & Americas\\
  46 & JMD   & Jamaican Dollar    & Frontier & Americas \\
  47 & PYG  & Paraguay Guarani    & Frontier & Americas \\
  48 & TTD  & Trinidad Tobago Dollar    & Frontier & Americas \\
  49 & HRK    & Croatian Kuna   & Frontier & Europe and CIS \\
  50 & KZT     & Kazakhstan Tenge    & Frontier & Europe and CIS \\
\hline \hline
\end{tabular}      
\label{ss:table1}
\end{table}

{\small
\begin{tabular}{clrrr}\hline\hline
 $ $ &  &  &  &  \\ 
 $i$ & Currency code & Currency name  & ~Type of economy & Geographical region \\ 
 \hline
  51 & LVL  & Latvian Lats     & Frontier & Europe and CIS\\ 
  52 & BWP  & Botswana Pula    & Frontier & Middle-East and Africa\\ 
  53 & KMF & Comoros Franc   & Frontier & Middle-East and Africa\\ 
  54 & GMD  & Gambian Dalasi   & Frontier & Middle-East and Africa   \\ 
  55 & GHC & Ghanaian Cedi      &     Frontier & Middle-East and Africa\\ 
  56 & GNF  & Guinea Franc   &    Frontier & Middle-East and Africa  \\ 
  57 & KES  & Kenyan Shilling     &     Frontier & Middle-East and Africa  \\ 
  58 & KWD  & Kuwaiti Dinar  &  Frontier & Middle-East and Africa  \\ 
  59 & MWK  & Malawi Kwacha  &  Frontier & Middle-East and Africa  \\ 
  60 & MRO  & Mauritanian Ouguiya  &  Frontier & Middle-East and Africa  \\ 
  61 & MZM  & Mozambique Metical  &  Frontier & Middle-East and Africa  \\ 
  62 & NGN  & Nigerian Naira  &  Frontier & Middle-East and Africa  \\ 
  63 & STD  & Sao Tome and Principe Dobra  &  Frontier & Middle-East and Africa  \\ 
  64 & SYP  & Syrian Pound   &  Frontier & Middle-East and Africa  \\ 
  65 & ZMK  & Zambian Kwacha  &  Frontier & Middle-East and Africa  \\ 
  66 & JOD  & Jordanian Dinar   &  Frontier & Middle-East and Africa  \\ 
  67 & BND  & Brunei Dollar  &  Frontier & Asia  \\
  68 & BDT  & Bangladeshi Taka  &  Frontier & Asia  \\ 
  69 & KHR  & Cambodian Riel  &  Frontier & Asia \\
  70 & FJD & Fiji Dollar  &  Frontier & Asia  \\ 
  71 & PKR & Pakistan Rupee  &  Frontier & Asia  \\ 
  72 & WST  & Samoan Tala  &  Frontier & Asia  \\ 
  73 & LKP  & Lao Kip & Frontier & Asia  \\ 
  74 & LKR  & Sri Lankan Rupee  &  Frontier & Asia  \\  
\hline \hline
\end{tabular}      
}

\section{The Return Cross-Correlation Matrix}
To quantify the degree of correlation between the exchange rate
movements for different
currencies, we first measure the fluctuations using the logarithmic
return so that the
result is independent of the scale of  measurement.
If $P_{i}(t)$ is the exchange rate of the $i$-th currency at time $t$
(in terms of USD), then the
logarithmic return is defined as
\begin{equation}
R_{i}(t,\Delta t) \equiv \ln {P_{i}(t+\Delta t)}- \ln {P_{i}(t)}.
\label{return}
\end{equation}
For daily return, $\Delta t$ = 1 day.
By dividing the time-series of returns thus obtained with their
standard deviation (which is a measure of the
volatility 
of the currency exchange rate),
$ \sigma_{i} = \sqrt{\langle R_{i}^{2} \rangle - \langle R_{i}
\rangle^{2}}$,
we obtain the normalized return,
$r_{i}(t,\Delta t) \equiv R_{i}/\sigma_{i}$. 
We observed that the cumulative distribution of the returns
displayed power-law scaling in the tails, i.e., $P (r_i > x) \sim
x^{- \alpha}$ where $\alpha$ is the corresponding exponent value. 
Using maximum likelihood
estimation, the exponents
for the different currencies were obtained and they were found to
be distributed over a narrow range of values with a peak
around $\alpha \simeq 3$. This indicates that the so-called {\em
inverse-cubic law} distribution of returns, reported in many studies
of stock price fluctuations~\cite{Lux96, Gopikrishnan98,Pan07b,Pan08}, 
also holds for currency exchange rate
movements~\cite{Koedijk90,Muller90}.
This further strengthens the {\em universality} of this empirical fact
about the nature of market fluctuations and supports the validity of explaining
this feature using very general models which do not consider details
of particular markets or economies (see, e.g., Ref.~\cite{Vikram11b}).

\begin{figure}%
\centering
\includegraphics[width=0.9\linewidth]{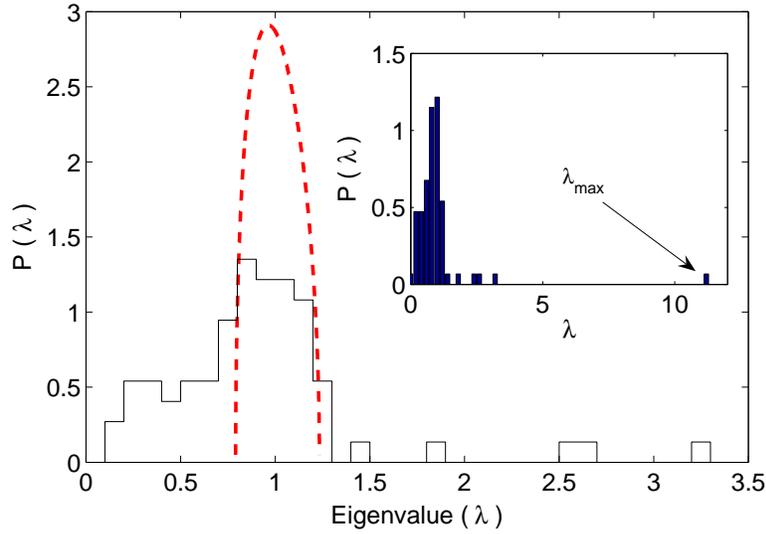}
\caption{The probability density function of the eigenvalues
of the cross-correlation matrix {\bf C} for fluctuations in the exchange rate in terms of
US Dollars of 74 currencies for the period Oct 1995-April 2012. For comparison the theoretical
distribution predicted by Eq.~(\ref{eq:sengupta}) is shown using broken curves. We explicitly
verified that the theoretical distribution fits very well the spectral distribution of
surrogate correlation matrices generated by randomly shuffling the returns for the different
currencies. The inset shows the largest eigenvalue corresponding to the global mode
of market dynamics.}%
\label{fig:eigv_dist}%
\end{figure}

After obtaining the 
return time series for all $N$ currencies over the period of $T$ days,
we calculate the cross-correlation matrix ${\mathbf C}$ whose
individual elements
$C_{ij}=\langle r_{i} r_{j} \rangle$,
represent the correlation between returns for a pair of currencies $i$ and $j$.
If the fluctuations of the different currencies are uncorrelated, the resulting random
correlation
matrix (referred to as a Wishart matrix) has eigenvalues distributed
according to~\cite{Sengupta99}:
\begin{equation}
P ( \lambda ) = \frac{Q}{2 \pi} 
\frac{\sqrt{(\lambda_{max} - \lambda)(\lambda -
\lambda_{min})}}{\lambda},
\label{eq:sengupta}
\end{equation}
with $N \rightarrow \infty$, $T \rightarrow \infty $ such that
$Q = T / N \geq 1$.
The bounds of the distribution are given by
$\lambda_{max} = [1 + (1/\sqrt{Q})]^2$ and $\lambda_{min} = 
[1 - (1/\sqrt{Q})]^2$. For the data we have analyzed, $Q = 81.54$, which implies
that in the absence of any correlation the spectral distribution should be bounded between
$\lambda_{max} = 1.23$ and $\lambda_{min} = 0.79$.
We observe from Fig.~\ref{fig:eigv_dist} that the bulk of the empirical eigenvalue
distribution indeed falls below the upper bound given by
$\lambda_{max}$, although a significant fraction of the
eigenvalues are smaller than what we expect from the lower bound
$\lambda_{min}$.
Also, a small number ($\simeq 8$) of the largest eigenvalues
are seen to deviate from the bulk of the distribution predicted by random matrix 
theory, and we focus our analysis on these modes
to obtain an understanding of the interaction structure of the world
currency market.

\begin{figure}%
\centering
\includegraphics[width=0.9\linewidth]{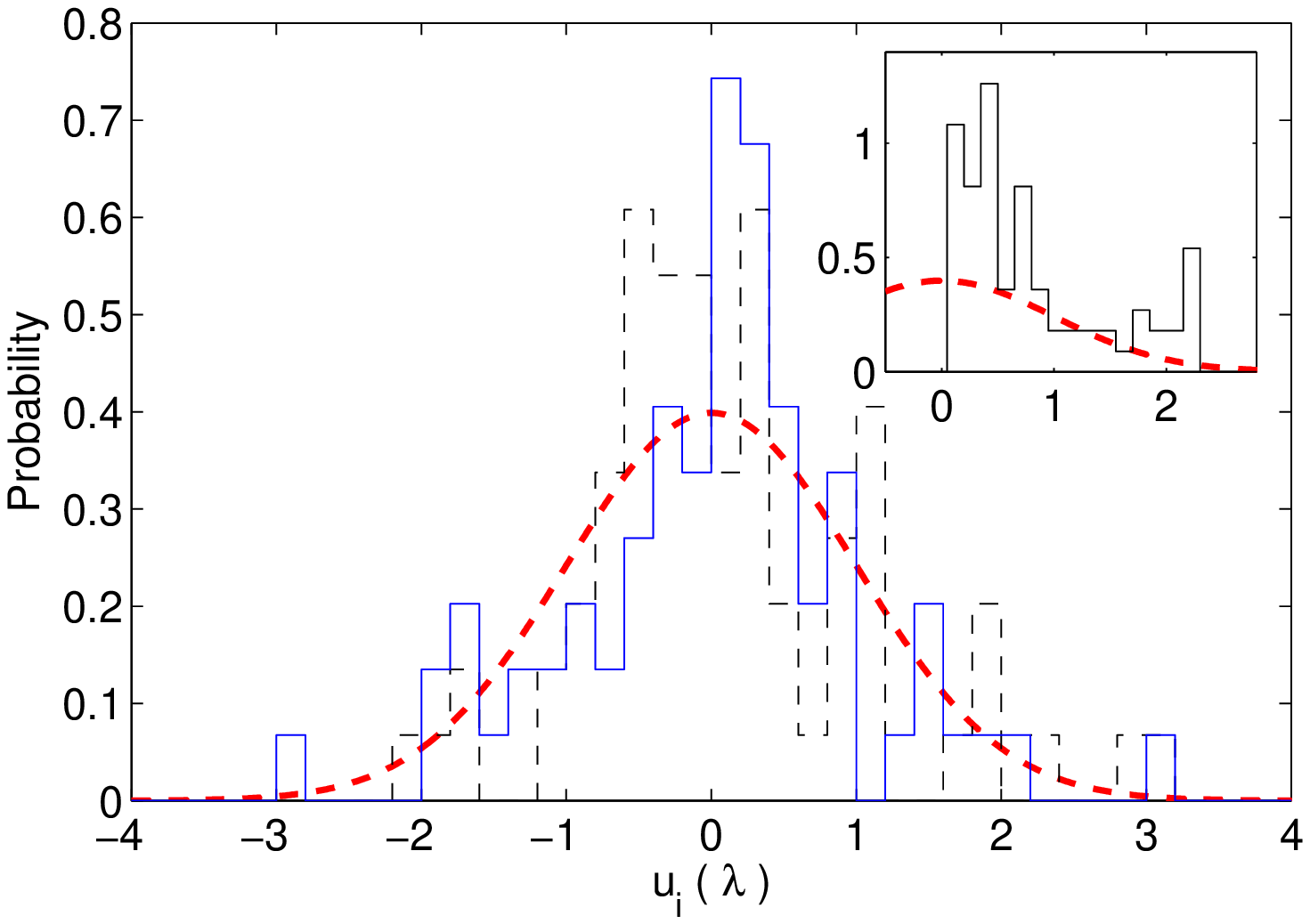}
\caption{The probability distribution of the eigenvector components corresponding to two
eigenvalues belonging to the bulk of the spectral distribution predicted by random matrix theory
and (inset) that corresponding to the largest eigenvalue. In both cases, the corresponding distribution
obtained from the surrogate correlation matrices obtained by randomly shuffling the returns is
shown using dotted lines for comparison.}%
\label{fig:eigvr_dist}%
\end{figure}

The random nature of the eigenvalues occurring in the bulk of the
distribution
is also indicated by the distribution of the corresponding eigenvector
components.
Note that, these components
are normalized for each eigenvalue $\lambda_{j}$ such that,
$\sum_{i=1}^{N}[u_{ji}]^2=N$, where $u_{ji}$ is the $i$-th
component of the $j$th eigenvector. 
For random matrices generated from uncorrelated time series, the
distribution of the eigenvector components follows the Porter-Thomas
distribution, 
\begin{equation} 
P(u)=\frac{1}{\sqrt{2\pi}}\exp[-\frac{u^2}{2}].
\label{ss:pt}
\end{equation}
We have explicitly verified this form for the corresponding
distribution of the random surrogate matrices obtained by
shuffling the empirical return time series so that all correlations
between the different currencies are destroyed.
As seen from Fig.~\ref{fig:eigvr_dist}, it also approximately fits
the distributions of the eigenvector components for the
eigenvalues
belonging to the bulk of the empirical spectral distribution. 
However, the eigenvectors of the largest
eigenvalues
(e.g., the largest eigenvalue $\lambda_{max}$, as shown in the inset)
deviate
quite significantly, indicating its non-random nature.

\begin{figure}[tbp]
\centering
\includegraphics[width=0.99\linewidth]{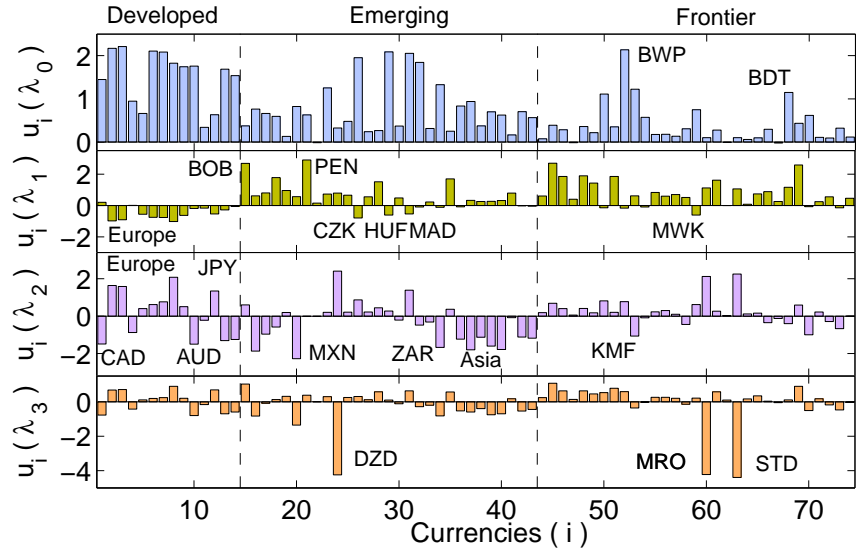}
\caption{The eigenvector components $u_i ( \lambda )$ for the four largest eigenvalues of
the correlation matrix {\bf C}. The currencies are arranged according
to the market classification of the
corresponding country (developed, emerging or frontier) separated by {\em broken lines}.
Some of the prominent components for each eigenvector (discussed in the text) 
are individually identified by the respective currency codes.}%
\label{fig:egvr_comp}%
\end{figure}

The largest eigenvalue $\lambda_0$ for the cross-correlation
matrix is about 9 times
larger than the upper bound of the random spectral distribution.
While this is similar to the situation for cross-correlations of stock
movements in financial markets (e.g., see Ref.~\cite{Sinha07,Pan07a}), the
corresponding eigenvector does not show a relatively uniform composition
unlike the case in equities markets where almost all stocks contribute to this
mode with all elements having the same sign. Instead, there is large
variation in the relative contributions of the different components to
the largest eigenmode, with those of four currencies (VEB, PYG, NGN,
BND) having a different
sign than the rest - although with an extremely low magnitude (Fig.~\ref{fig:egvr_comp},
top).
This eigenmode represents the global component of the time-series of currency
fluctuations which is common to all currencies. Thus, the strength of
the relative contribution of a currency to the leading eigenvector can
be construed as the extent to which the currency is in sync with the
the overall movement of the world currency market reflecting the collective response 
of the world economy to information shocks (which may include major
perturbations such as the worldwide
financial crisis of 2007-08).
Note that, this suggests that the relative strengths of the components
in the leading eigenvector may be used as a measure of the role
the corresponding currency plays in the world market (and to an extent,
that the country plays in
the international economy). Seen from this point of view, it is
perhaps not surprising that most of the currencies belonging to
countries in the developed markets category contribute significantly
to this mode which reflects their dominance in the world economy.
We also see that the countries in the emerging 
markets category can be very different from each other in terms of
their role in the global mode, with components corresponding to the
East European economies such as Czech Republic, Hungary and Poland having 
some of the largest contributions.
Turning to the frontier markets category, while the contributions of most 
of these currencies have very low magnitude, a few countries (most
notably Botswana but also Bangladesh, Kazakhstan and Comoros) 
stand out for the relatively high strength of the
corresponding eigenvector component. 
The strong contribution from these countries could be either because
of their impressive economic performance (e.g., Botswana has
maintained one of the world's highest economic growth rates from the
time of its independence in 1966~\cite{cia}) or possibly due to remittances in
foreign currencies from expatriates working abroad having a large
contribution to the national economy (as in the case of Bangladesh).
As newly developing economies are 
potentially highly profitable but risky targets for foreign investment,
it may be of interest to explore the possibility of using this measure
to identify frontier markets having strong interaction with the world market which
may make them relatively safer to invest in. On the other hand, from
the point of view of portfolio diversification for reducing risk, one may use such a
measure to identify economies whose fluctuations have the least in
common with the global mode.

Of even more interest for understanding the topological structure of
interactions in the world currency market are the
intermediate
eigenvalues in between the largest eigenvalue $\lambda_0$ and the bulk
predicted by random matrix theory. For equities markets, it has been
shown that in many cases the eigenvectors corresponding to these
eigenvalues
are localized, i.e., a relatively small number of stocks, usually
having similar market capitalization or belonging to the same
business sector, contribute
significantly to these modes~\cite{Pan07a,Gopikrishnan01,Plerou02}. 
Fig.~\ref{fig:egvr_comp} shows that the different currencies
contribute to the different eigenvectors corresponding to the three
largest intermediate eigenvalues
very unequally. For example, from the eigenvector corresponding to
$\lambda_1$, the second
largest eigenvalue, we observe that many Latin American currencies such
as those of Bolivia and Peru, have a
dominant contribution in this mode with the contribution of European 
currencies (and a few non-European ones, such as those of Morocco and Malawi,
where the corresponding 
economy is closely connected to that of Europe) being not only
different but actually having the opposite sign.
The third eigenvector exhibits contributions of different signs from European and
Japanese currencies on the one hand, and established as well as rapidly
developing economies of America, Asia-Pacific and Africa (such as Canada,
Mexico, South Africa, Australia, New Zealand, Israel, Singapore and India)
on the other. The fourth eigenvector has significant contributions
from only three currencies, those of Algeria, Mauritania and Sao Tome
\& Principe. This may reflect existing economic linkages between these
countries that has resulted in such strong coupling in the movements
of their currency exchange rates with respect to the US Dollar.

\begin{figure}[tbp]
\centering
\includegraphics[width=0.9\linewidth]{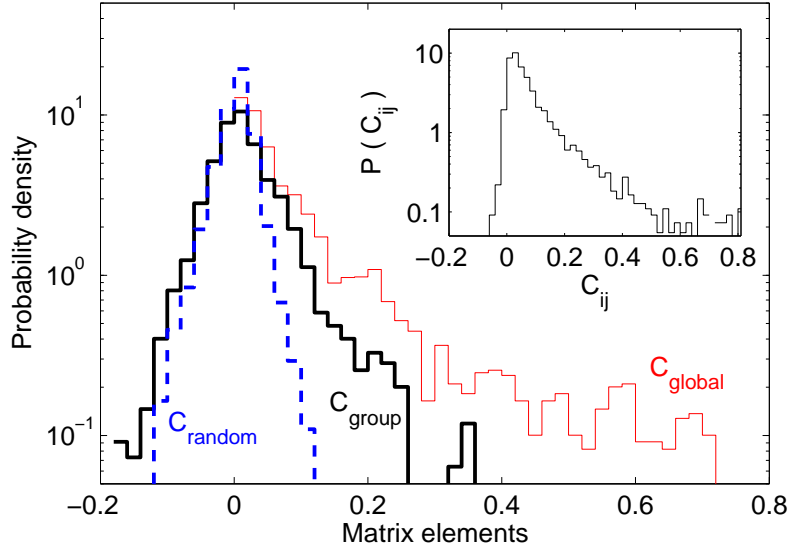}
\caption{The probability distribution of the matrix elements following decomposition of the
correlation matrix {\bf C} into global ({\bf C}$_{global}$), group ({\bf C}$_{group}$) and random
effects ({\bf C}$_{effects}$) with $N_{g} = 7$. The distribution of the components $C_{ij}$ of the
original cross-correlation matrix {\bf C} is shown in the inset for comparison.
}%
\label{fig:corr}%
\end{figure}

Despite the above insights, a direct inspection of eigenvector composition for the
intermediate eigenvalues does not very often yield a straightforward
interpretation of the group of currencies dominantly contributing to a
particular mode. This is because
apart from information about interactions between
currencies, the cross-correlations are also affected strongly by the global mode 
corresponding to the overall market movement. In addition, there are a 
large number of modes belonging to the random bulk
which correspond to idiosyncratic fluctuations. Both the global and
random modes can mask
significant intra-group correlations.
Thus, in order to identify the topological structure of
interactions
between the currencies we need to remove the global mode corresponding
to the largest eigenvalue and also filter out the effect of random
noise (contributed by the eigenvalues belonging to the bulk of the spectral distribution).
For this we use the filtering method proposed in
Ref.~\cite{Kim05} based on the expansion of a
matrix in terms of its eigenvalues
$\lambda_{i}$ and the corresponding eigenvectors ${\mathbf u}_{i}$:
${\mathbf C}= \Sigma_i \lambda_i \mathbf{u}_i \mathbf{u}_i^T$. This
allows the correlation matrix
to be decomposed into three parts, corresponding to the global, group
and random components:
\begin{equation}
{\mathbf C} = {\mathbf C}_{global} + {\mathbf C}_{group} +
{\mathbf C}_{random} = \lambda_0 \mathbf{u}_0^T \mathbf{u}_0 +
\sum_{i =1}^{N_{g}} \lambda_i \mathbf{u}_i^T \mathbf{u}_i +
\sum_{i = N_{g}+1}^{N-1} \lambda_i \mathbf{u}_i^T \mathbf{u}_i,
\end{equation}
where, the eigenvalues have been arranged in descending order (the
largest
labelled 0) and $N_{g}$ is the number of intermediate eigenvalues.
From the empirical data it may not be obvious what is the value of
$N_{g}$, as the bulk may differ from the predictions of random matrix
theory because of underlying structure induced correlations. For this
reason, we use visual inspection to choose $N_{g}
= 6$, and verify that small changes in this value do not alter the
results.
Our results are robust with respect to small variations in the estimation of
$N_{g}$ because the error involved is only due to the
eigenvalues closest to the bulk that have the smallest contribution to
${\mathbf C}_{group}$. Fig.~\ref{fig:corr}
shows the result of the decomposition
of the entire cross-correlation matrix (shown in the inset) into the three components.
In contrast to the case of stock-stock correlations in financial
markets (e.g., Ref.~\cite{Pan07a}), in the currency market
the group correlation matrix elements $C_{ij}^{group}$ show a
significantly reduced tail and is completely enveloped by 
the distribution of the global correlation
matrix elements $C_{ij}^{global}$. This implies that there are a
relatively small fraction of strongly interacting currencies, 
implying that the segregation into groups may be weak in this market.

\begin{figure}[tbp]
\centering
\includegraphics[width=0.99\linewidth]{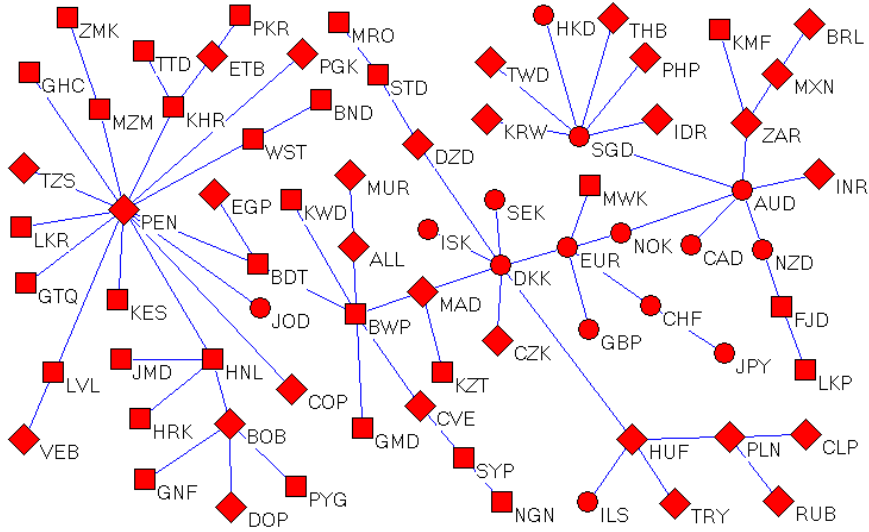}
\caption{The minimum spanning tree connecting the 74 currencies considered here. The node shapes indicate the
type of the underlying economy of the country to which the currency belongs (circles indicate developed,
diamonds indicates emerging and squares indicate frontier markets). The figure has been drawn using
the Pajek software.
}%
\label{fig:mst}%
\end{figure}

In order to graphically present the interaction structure of the
stocks using the information in the group correlation matrix {\bf
C}$_{group}$, we first 
use a method suggested by Mantegna~\cite{Mantegna99} to
transform
the correlation between currencies into distances to produce a connected
network in which co-moving currencies are clustered together.
The distance $d_{ij}$ between two currencies $i$ and $j$ are calculated
from the
cross-correlation matrix ${\mathbf C}$, according to
$d_{ij} = \sqrt{2 (1 - C_{ij})}$. These are used to construct a
minimum
spanning tree, which connects
all the $N$ nodes of a network with $N-1$ edges such that the
total sum of the distance between every pair of nodes, $\sum_{i,j}
d_{ij}$, is
minimum. 
As seen in Fig.~\ref{fig:mst}, for the currency market this method
reveals clusters of currencies belonging to countries having
similar economic profile and/or belonging to the same geographical
region. In particular, note the cluster centered around the hub node
(i.e., a node having significantly more connections than the average)
corresponding to SGD which consists exclusively of currencies
belonging to developed or emerging economies of the Asia-Pacific region
such as those of Hong Kong, Taiwan, Thailand, Indonesia etc. On the other
hand, the currencies clustered around the hub AUD are related by the
geo-economic status of the corresponding countries of being 
major non-European players in the world economy 
(e.g., Canada, Mexico, Brazil, South Africa and India).
It should be noted that the hubs of these two clusters (SGD and
AUD) are directly linked to each other and are in turn
connected to the cluster of European currencies (comprising two
hubs corresponding to the Euro and the Danish currency) suggesting 
a close interplay in the currency movements of all the important
countries driving international economic dynamics. 
Possibly more intriguing is the occurrence of a much bigger cluster
(containing a third of all the currencies considered) 
arranged around the largest hub in the network which corresponds 
to the Peruvian currency. 
This cluster comprises a wide assortment of currencies belonging to countries
spread geographically around the world but which share an economic resemblance
in that most of them are in a relative state of underdevelopment
compared to the economies considered earlier.
It thus appears that the tree network representing the underlying interactions 
in the world currency market can be approximately divided into a part
comprising developed or rapidly growing economies (dominated by Europe and
Asia-Pacific) and another part composed of relatively underdeveloped ones
(consisting mostly of Latin American and African countries), with
the currency movements of these two groups being relatively independent of each
other. Note that the two parts, in particular, the hubs corresponding to PEN and DKK,
are bridged by the currencies of Morocco,
Botswana and Bangladesh, which therefore have an importance in governing
the collective dynamics of the world economy disproportionate to their intrinsic economic
status. This can potentially explain the strong contribution of these currencies
to the leading eigenvector of the cross-correlation matrix that represents the
global eigenmode which has been discussed earlier in this article.
     
\begin{figure}[tbp]
\centering
\includegraphics[width=0.99\linewidth]{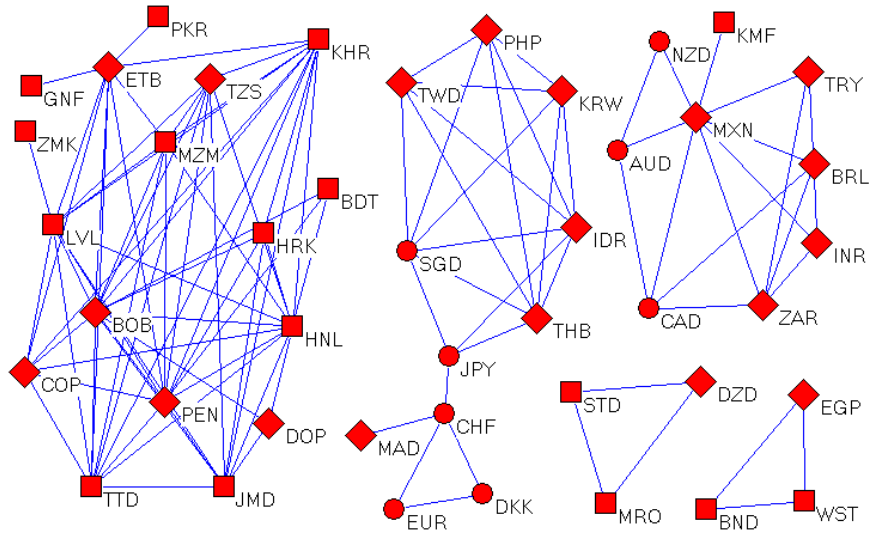}
\caption{The network of interactions among currencies generated from the group correlation matrix
{\bf C}$_{group}$ with threshold $c^{*} = 0.133$. The node shapes indicate the
type of the underlying economy of the country to which the currency belongs (circles indicate developed,
diamonds indicates emerging and squares indicate frontier markets). The cluster at the center consists mostly
of countries belonging to the Asia-Pacific region including several
members of the ASEAN group, although it is also connected via the
Japanese Yen to a smaller sub-group of European currencies. The cluster at top right
consists of three of the BRICS countries
as well as several economies outside
Europe which are important in the global economy (such as Australia, Canada, Mexico and Turkey).
The cluster at the left comprises mostly Latin American and African currencies - although note the
presence of Bangladesh and Brunei. The two small clusters at the bottom connect triads of currencies.
The figure has been drawn using
the Pajek software.
}%
\label{fig:clusters}%
\end{figure}

We have also used an alternative method of graph visualization in order to
highlight any existing groups of currencies having significant mutual
interactions. For
the case of stocks in financial markets, the modules obtained by
this technique often represent strongly performing business sectors in
the economy~\cite{Sinha07,Pan07a}. It is thus plausible that the
currency communities identified using this method will represent 
important groupings driving the world economy.
The binary-valued adjacency
matrix ${\mathbf A}$ of the network is generated from
${\mathbf C}_{group}$ by using a threshold $c_{th}$ such that $A_{ij}
= 1$
if $C_{ij}^{sector} > c_{th}$, $A_{ij} = 0$ otherwise. An appropriate
choice of the threshold makes apparent any clustering in the network
that is implied by the existence of a 
tail in the
$C_{ij}^{group}$ distribution.
Fig.~\ref{fig:clusters} shows the resultant network for the best choice
of
$c_{th}=c^*$ (= 0.133) in terms of creating the largest clusters of
interacting currencies (isolated nodes have not been shown).
The five clusters differ considerably in size, with two of them
corresponding to strongly interacting currency triads (with the
DZD-MRO-STD triad being the currencies having the dominant
contribution to the fourth largest eigenmode identified earlier in
Fig.~\ref{fig:egvr_comp}). The next largest cluster, having
nine currencies, consists of rapidly emerging economies outside Europe - including
Brazil, India and South Africa of the BRICS group as well as Turkey
and Mexico from the ``Next Eleven" (N-11) group identified in
Ref.~\cite{Oneill05}
as countries having potential of becoming some of the largest
economies in the world in the coming years - and a few
non-European developed economies such as Australia and Canada.
The even larger cluster comprising eleven currencies is dominated by
the countries of Asia-Pacific such as Taiwan and Singapore as well 
as the N-11 countries Indonesia, Korea and Philippines, which have
either developed or fast growing economies; however, through the Japanese
Yen, these currencies are also connected to a smaller sub-cluster of
European currencies which contains the Euro apart from the Swiss and
Danish currencies (note also the presence of the currency of Morocco,
a north African country but one that has strong economic ties with Europe).
The largest cluster has seventeen densely inter-connected currencies
which are geographically spread around the world, although half of
them are from Latin America or the Caribbean. Possibly this cluster
reflects a new wave of fast growing economies (e.g., it includes
two N-11 countries, Bangladesh and Pakistan) whose development
trajectory may affect the global economy in the future as profoundly as the rise of
India and China has affected it in the past decades.

\section{Conclusions}
In this article we have analyzed the topological structure of interactions
in the world currency market by using the spectral properties of the cross-correlation
matrix of exchange rate fluctuations. We see that the eigenvalue
distribution is similar to that seen in equities markets and consists of a bulk approximately
matching the predictions of random matrix theory. In addition, there
are several deviating
eigenvalues which contain important information about groups of strongly
interacting components.
However, the composition of the leading eigenvector shows a remarkable
distinction in that, unlike the relatively homogeneous nature of the
eigenvector for cross-correlations in the equities market where all
stocks contribute almost equally to the market or global mode, the
different currencies can have widely differing contributions to the
global mode for exchange rate cross-correlations. This possibly reflects
the extent to which the fluctuations of a currency is in sync with the overall
market movement and may also be used to measure the influence of a
currency in the world economy. While, as is probably expected, the large components
of this mode mostly belong to currencies
of the developed economies of western Europe as well as the rapidly growing
economies of the Asia-Pacific region, there are unexpectedly strong
contributions
from currencies outside this group - such as those of Botswana, Bangladesh 
and Kazakhstan. This indicates that these economies may be playing an important
role in directing the collective dynamics of the international currency market that
is not exclusively dependent on their intrinsic economic strength, but rather
the position they occupy in the network of interactions among the currencies. 
This is confirmed by the reconstructed network of interactions among the currencies
as a minimum spanning tree. This network shows a segregation between clusters dominated
by developed or rapidly growing economies on the one hand, and relatively
underdeveloped economies on the other. While these two parts can show dynamics
relatively independent of each other, a few currencies - those of Morocco, Botswana
and Bangladesh - act as a bridge between them. Thus the role of these currencies as 
vital connecting nodes of the world currency market possibly give them a
much more important position than would be expected otherwise.
We have also used an alternative graph representation technique to identify
several groups of strongly interacting currencies. Some of the smaller clusters
may be reflecting possible economic or other relations between the corresponding
countries. However, the largest cluster comprises a densely interconnected set
of currencies belonging to 
countries that are geographically spread apart. We speculate that these could
well belong to the next wave of fast emerging economies that will drive 
the economic growth of the world in the future. This is significant 
from the point of view of applications, as such economies are potentially lucrative
targets for foreign investment and are eagerly sought after by portfolio
fund managers. Methods of identifying early the next fast growth economies
assume critical importance in such a situation. Our analysis of cross-correlations
of exchange rate fluctuations suggests that prominent clusters in the reconstructed 
networks of interactions in the world currency market may potentially provide
us with such methods.   

\begin{acknowledgement}
We would like to thank R K Pan who helped in developing the software 
used for the analysis reported here and S Sridhar for stimulating
discussions. Part of the work was supported by the Department of
Atomic Energy, Government of India through the IMSc Complex
Systems Project (XII Plan).
\end{acknowledgement}

\end{document}